\documentclass[showpacs,twocolumn,preprintnumbers,amsmath,amssymb]{revtex4}

\usepackage{graphicx}
\usepackage{dcolumn}
\usepackage{bm}

\newcommand{\bq}{\begin{equation}}
\newcommand{\eq}{\end{equation}}
\newcommand{\bqa}{\begin{eqnarray}}
\newcommand{\eqa}{\end{eqnarray}}
\newcommand{\nn}{\nonumber \\}

\def\be     {\begin{equation}}
\def\ee     {\end{equation}}
\def\bea        {\begin{eqnarray}}
\def\eea        {\end{eqnarray}}
\def\bnn    {\begin{eqnarray*}}
\def\enn    {\end{eqnarray*}}

\begin{document}

\title{Role of axion electrodynamics in Weyl metal: Violation of Wiedemann-Franz law}
\author{Ki-Seok Kim}
\affiliation{ $^{1}$Department of Physics, POSTECH, Pohang, Gyeongbuk 790-784, Korea \\ $^{2}$Institute of Edge of
Theoretical Science (IES), Hogil Kim Memorial building 5th floor, POSTECH, Pohang, Gyeongbuk 790-784, Korea }
\date{\today}

\begin{abstract}
Recently, enhancement of the longitudinal magneto-electrical conductivity (LMEC) has been observed in Bi$_{1-x}$Sb$_{x}$ around $x \sim 3\%$ under $\bm{E} \parallel \bm{B}$ ($\bm{E}$: external electric field and $\bm{B}$: external magnetic field) [Phys. Rev. Lett. {\bf 111}, 246603 (2013)], where an enhancement factor proportional to $B^{2}$ is suggested to result from the $\bm{E} \cdot \bm{B}$ term. In the present study, we show that this $B^{2}$ enhancement is not limited on the LMEC, where both the Seebeck and thermal conductivities in the longitudinal setup ($\bm{E} \parallel \bm{B}$) are predicted to show essentially the same enhancement proportional to $B^{2}$. In particular, the $B^{2}$ enhancement factor of the LMEC turns out to differ from that of the longitudinal thermal conductivity, responsible for breakdown of Wiedemann-Franz (WF) law, which means that anomalous currents flowing through the dissipationless channel differ from each other. Since the breakdown of the WF law appears in spite of the existence of electron quasiparticles, regarded to be a purely topological character (chiral anomaly), the Weyl metallic state cannot be identified with the Landau's Fermi-liquid fixed point. We propose the violation of the WF law as another hallmark of the Weyl metallic phase, which originates from axion electrodynamics.
\end{abstract}

\maketitle

Weyl metal \cite{Weyl_Metal1,Weyl_Metal2,Weyl_Metal3} is a topological Fermi-liquid state with a pair of chiral Fermi surfaces, which encode not only Berry curvature but also chiral anomaly \cite{Haldane,Kim_Review}. As a result, descriptions based on conventional Maxwell equations fail to detect their electromagnetic properties. Instead, their characteristic features are described by axion electrodynamics, given by the topological-in-origin $\bm{E}\cdot\bm{B}$ term \cite{Axion_EM}. Then, it is natural to investigate how well-known physical properties of a non-topological Fermi-liquid state will be modified by the axion electrodynamics of a pair of chiral Fermi surfaces.

The first step would be to construct an effective field theory for such a topological metallic state. An effective action functional for single particle dynamics has been constructed in the phase pace, imposing the momentum-space Berry gauge-field \cite{Chiral_Magnetic_Effect1,Chiral_Magnetic_Effect2,CEFT_QED4_1,CEFT_QED4_2}. Reformulating the Berry phase based on the Stoke's theorem, a five-dimensional gauged Wess-Zumino-Witten (WZW) term has been proposed to reproduce chiral anomalies associated with not only electromagnetic but also gravitational fields and their mixing \cite{TLFL_EFT1,TLFL_EFT2,TLFL_EFT3}. Unfortunately, this fascinating construction does not seem to be much practical for actual calculations because it is not easy to introduce the role of such a term into diagrammatic analysis as the case of field theories with topological terms. A more practical version of an effective field theory needs to be constructed.

A Boltzmann-equation approach has been constructed to deduce anomalous electromagnetic transport phenomena such as chiral magnetic effect \cite{Chiral_Magnetic_Effect1,Chiral_Magnetic_Effect2,Chiral_Magnetic_Effect3,Chiral_Magnetic_Effect4,Chiral_Magnetic_Effect5,Chiral_Magnetic_Effect6,Chiral_Magnetic_Effect7,Chiral_Magnetic_Effect8} and negative longitudinal magnetoresistivity \cite{Negative_LMR1,Negative_LMR2,Negative_LMR3}, where effects of both Berry curvature and chiral anomaly, i.e., axion electrodynamics are incorporated by semiclassical equations of motion. Recently, this phenomenological Boltzmann-equation framework has been derived from QED$_{4}$ (four-dimensional quantum electrodynamics) with an axion term, resorting to the gauge-invariant Wigner distribution function \cite{QBE_WM1,QBE_WM2}. At present, a transport theory based on the Boltzmann-equation approach is the reality of a topological Fermi-liquid theory \cite{Haldane,Kim_Review}.

In this paper, we investigate both thermal and thermoelectric transport properties of Weyl metal based on the Boltzmann-equation approach. A more natural and consistent description for thermal transport properties would be to extend the Boltzmann-equation framework in a general covariant way, where not only the Boltzmann equation itself but also semiclassical equations of motion are constructed on a curved space and time. Here, we follow the conventional approach, introducing temperature-gradients into the Boltzmann equation explicitly, where semiclassical equations of motion are unchanged to keep the information of Berry curvature and chiral anomaly on electromagnetism. 
%
%

We start from transport equations, given by
\bqa && J_{i}^{e} = L_{ij}^{ee} E_{j} + L_{ij}^{eq} \Bigl( - \frac{\partial_{j} T}{T} \Bigr) , \nn && J_{i}^{q} = L_{ij}^{qe} E_{j} + L_{ij}^{qq} \Bigl( - \frac{\partial_{j} T}{T} \Bigr) , \eqa where $J_{i}^{e}$ and $J_{i}^{q}$ are electric and thermal currents flowing into the $i-$direction and $E_{j}$ and $\partial_{j} T$ are applied electric fields \cite{Comment_on_E} and temperature gradients of the $j-$direction, respectively. Transport coefficients are
\bqa && \sigma_{ij} = L_{ij}^{ee} , ~~~ s_{ij} = - \frac{1}{T} [\bm{L}^{ee}]^{-1}_{ik} L_{kj}^{eq} , \nn && \kappa_{ij} = \frac{1}{T} \Bigl( L_{ij}^{qq} - L_{ik}^{qe} [\bm{L}^{ee}]^{-1}_{kl} L_{lj}^{eq} \Bigr) , \eqa where $\sigma_{ij}$ is an electrical conductivity tensor with an index of $ij$, $s_{ij}$ is the Seebeck coefficient tensor, and $\kappa_{ij}$ is a thermal conductivity tensor.

An idea is to incorporate the information of the topological structure of Weyl metal into the Boltzmann-equation approach via semiclassical equations of motion with Berry curvature, which detects semiclassical dynamics of a particle on a chiral Fermi surface \cite{Semiclassical_Eqs1,Semiclassical_Eqs2}. The existence of Berry curvature gives rise to the following set of solutions 
\begin{widetext}
\bqa && \boldsymbol{\dot{r}}_{\pm} = \Bigl( 1 + \frac{e}{c} \boldsymbol{B} \cdot \boldsymbol{\Omega}_{\boldsymbol{p}}^{\pm} \Bigr)^{-1} \Bigl\{ \boldsymbol{v}_{\boldsymbol{p}} + e \boldsymbol{E} \times \boldsymbol{\Omega}_{\boldsymbol{p}}^{\pm} + \frac{e}{c} \boldsymbol{\Omega}_{\boldsymbol{p}}^{\pm} \cdot \boldsymbol{v}_{\boldsymbol{p}} \boldsymbol{B} \Bigr\} , \nn && \boldsymbol{\dot{p}}_{\pm} = \Bigl( 1 + \frac{e}{c} \boldsymbol{B} \cdot \boldsymbol{\Omega}_{\boldsymbol{p}}^{\pm} \Bigr)^{-1} \Bigl\{ e \boldsymbol{E} + \frac{e}{c} \boldsymbol{v}_{\boldsymbol{p}} \times \boldsymbol{B} + \frac{e^{2}}{c} (\boldsymbol{E} \cdot \boldsymbol{B}) \boldsymbol{\Omega}_{\boldsymbol{p}}^{\pm} \Bigr\} . \eqa 
\end{widetext} 
$\bm{B}$ and $\bm{E}$ are applied magnetic and electric fields, respectively. The momentum-space Berry curvature is given by $\bm{\nabla}_{\bm{p}} \cdot \bm{\Omega}_{\bm{p}}^{\pm} = \pm \delta^{(3)}(\bm{p} \pm g \bm{B})$, where the superscript $+(-)$ means a positive (negative) ``magnetic" charge corresponding to a $+(-)$ chirality Weyl point, separated by the applied magnetic field with the Land$\acute{e}-g$ factor \cite{KSKim_BE}. In addition to well-known terms of group velocity and Lorentz force, there appear three unconventional terms. $e \boldsymbol{E} \times \boldsymbol{\Omega}_{\boldsymbol{p}}^{\pm}$ is referred to as an anomalous velocity, responsible for anomalous Hall effect \cite{Semiclassical_Eqs1,Semiclassical_Eqs2}. This term results from Berry curvature purely, allowed even in non-chiral Fermi surfaces \cite{Haldane}. On the other hand, $\frac{e}{c} \boldsymbol{\Omega}_{\boldsymbol{p}}^{\pm} \cdot \boldsymbol{v}_{\boldsymbol{p}} \boldsymbol{B}$ and $\frac{e^{2}}{c} (\boldsymbol{E} \cdot \boldsymbol{B}) \boldsymbol{\Omega}_{\boldsymbol{p}}^{\pm}$ emerge only in chiral Fermi surfaces, responsible for chiral magnetic effect \cite{Chiral_Magnetic_Effect1,Chiral_Magnetic_Effect2,Chiral_Magnetic_Effect3,Chiral_Magnetic_Effect4,Chiral_Magnetic_Effect5,Chiral_Magnetic_Effect6,Chiral_Magnetic_Effect7,Chiral_Magnetic_Effect8} and negative longitudinal magnetoresistivity \cite{Negative_LMR1,Negative_LMR2,Negative_LMR3}, respectively.

Incorporating these solutions into the Boltzmann-equation framework, we reach an effective transport theory
\begin{widetext}
\bqa && \Bigl( 1 + \frac{e}{c} \boldsymbol{B} \cdot \boldsymbol{\Omega}^{\pm}_{\boldsymbol{p}} \Bigr)^{-1} \Bigl\{ \boldsymbol{v}_{\boldsymbol{p}} + \frac{e}{c} \boldsymbol{\Omega}^{\pm}_{\boldsymbol{p}} \cdot \boldsymbol{v}_{\boldsymbol{p}} \boldsymbol{B} \Bigr\} \cdot \Bigl( \bm{\nabla}_{\bm{r}} \mu^{\pm} + (\epsilon_{p} - \mu^{\pm}) \frac{\bm{\nabla}_{\bm{r}} T}{T} \Bigr) \Bigl( - \frac{\partial f_{eq}(\epsilon_{p})}{\partial \epsilon_{p}} \Bigr) \nn && + \Bigl( 1 + \frac{e}{c} \boldsymbol{B} \cdot \boldsymbol{\Omega}^{\pm}_{\boldsymbol{p}} \Bigr)^{-1} \Bigl\{ e \boldsymbol{E} + \frac{e}{c} \boldsymbol{v}_{\boldsymbol{p}} \times \boldsymbol{B} + \frac{e^{2}}{c} (\boldsymbol{E} \cdot \boldsymbol{B}) \boldsymbol{\Omega}^{\pm}_{\boldsymbol{p}} \Bigr\} \cdot \bm{\nabla}_{\bm{p}} f^{\pm}(\bm{p})
= - \Gamma_{imp} [f_{\pm}(\bm{p}) - f_{eq}(\bm{p})] - \Gamma'_{imp} [f_{\pm}(\bm{p}) - f_{\mp}(\bm{p})] , \nn \eqa
\end{widetext}
where $f_{\pm}(\bm{p})$ represent distribution functions on $\pm$ chiral Fermi surfaces with chemical potentials $\mu^{\pm}$, and $\Gamma_{imp}$ and $\Gamma_{imp}'$ are ``intra-valley" and ``inter-valley" scattering rates, respectively, phenomenologically introduced and $\Gamma_{imp}'/\Gamma_{imp} \ll 1$ assumed for simplicity. Electric and thermal currents are constructed as
\begin{widetext}
\bqa && \bm{j}_{el}^{\pm} = e \int \frac{d^{3} \bm{p}}{(2\pi)^{3}} \Bigl\{ \boldsymbol{v}_{\boldsymbol{p}} + e \boldsymbol{E} \times \boldsymbol{\Omega}_{\boldsymbol{p}}^{\pm} + \frac{e}{c} \boldsymbol{\Omega}_{\boldsymbol{p}}^{\pm} \cdot \boldsymbol{v}_{\boldsymbol{p}} \boldsymbol{B} \Bigr\} f_{\pm}(\bm{p}) , \nn && \bm{j}_{th}^{\pm} = \int \frac{d^{3} \bm{p}}{(2\pi)^{3}} (\epsilon_{p} - \mu^{\pm}) \Bigl\{ \boldsymbol{v}_{\boldsymbol{p}} + e \boldsymbol{E} \times \boldsymbol{\Omega}_{\boldsymbol{p}}^{\pm} + \frac{e}{c} \boldsymbol{\Omega}_{\boldsymbol{p}}^{\pm} \cdot \boldsymbol{v}_{\boldsymbol{p}} \boldsymbol{B} \Bigr\} f_{\pm}(\bm{p}) . \eqa
\end{widetext}
It is interesting to observe an analogous contribution of chiral magnetic effect in the heat current, also proportional to the distance of a pair of Weyl points.

It is straightforward to solve these coupled Boltzmann equations. We refer all details to our recent study on electrical transport properties in Weyl metal \cite{KSKim_BE}. Both electrical and thermal currents are found to be
\begin{widetext}
\bqa && j_{el}^{\pm x} \approx \frac{e^{2}}{c} \int \frac{d^{3} \bm{p}}{(2\pi)^{3}} (\boldsymbol{\Omega}_{\boldsymbol{p}}^{\pm} \cdot \boldsymbol{v}_{\boldsymbol{p}}) f_{eq}(\bm{p}) B \nn && + e \int \frac{d^{3} \bm{p}}{(2\pi)^{3}} \Bigl( - \frac{\partial f_{eq}(\epsilon_{p})}{\partial \epsilon_{p}} \Bigr) \Bigl\{ \Bigl( v_{x}(\bm{p}) + \frac{e}{c} B \boldsymbol{\Omega}_{\boldsymbol{p}}^{\pm} \cdot \bm{v}_{\bm{p}} \Bigr)^{2} + [v_{x}(\bm{p})]^{2} \frac{\Gamma_{imp}'}{\Gamma_{imp}} \Bigr\} \frac{1}{\Gamma_{imp}} \Bigl\{ e \Bigl( E - \frac{1}{e} \partial_{x} \mu^{\pm} \Bigr) - (\epsilon_{p} - \mu^{\pm}) \frac{\partial_{x} T}{T} \Bigr\} , \nn && j_{th}^{\pm x} \approx \frac{e}{c} \int \frac{d^{3} \bm{p}}{(2\pi)^{3}} (\boldsymbol{\Omega}_{\boldsymbol{p}}^{\pm} \cdot \boldsymbol{v}_{\boldsymbol{p}}) (\epsilon_{p} - \mu^{\pm}) f_{eq}(\bm{p}) B \nn && + \int \frac{d^{3} \bm{p}}{(2\pi)^{3}} \Bigl( - \frac{\partial f_{eq}(\epsilon_{p})}{\partial \epsilon_{p}} \Bigr) \Bigl\{ \Bigl( v_{x}(\bm{p}) + \frac{e}{c} B \boldsymbol{\Omega}_{\boldsymbol{p}}^{\pm} \cdot \bm{v}_{\bm{p}} \Bigr)^{2} + [v_{x}(\bm{p})]^{2} \frac{\Gamma_{imp}'}{\Gamma_{imp}} \Bigr\} \frac{1}{\Gamma_{imp}} (\epsilon_{p} - \mu^{\pm}) \Bigl\{ e \Bigl( E - \frac{1}{e} \partial_{x} \mu^{\pm} \Bigr) - (\epsilon_{p} - \mu^{\pm}) \frac{\partial_{x} T}{T}  \Bigr\} . \nn \eqa
\end{widetext}
An essential aspect is that the applied temperature gradient appears to be exactly the same way as the applied electric field. This observation leads us to conclude that not only electrical resistivity but also both thermal resistivity and Seebeck ``resistivity" will show the negative longitudinal magnetoresistivity \cite{Negative_LMR3,KSKim_BE}. In other words, such resistivities become suppressed when currents are driven along the direction of a pair of Weyl points since a dissipationless current channel is formed along that direction ``due to" the chiral anomaly.

It is not difficult to read transport coefficients from the above expression, given by
\begin{widetext}
\bqa &&  L_{xx}^{ee} = \sigma_{0} \Bigl\{ 1 + \frac{\Gamma_{imp}'}{\Gamma_{imp}} + \Bigl(\frac{e}{c}\Bigr)^{2} \Bigl( \int_{|\bm{p}| = p_{F}} d \mathcal{S}_{\bm{p}} [\boldsymbol{\Omega}_{\boldsymbol{p}} \cdot \bm{\hat{n}}_{\bm{p}}]^{2}  \Bigr) B^{2} \Bigr\} , \nn && L_{xx}^{eq} = L_{xx}^{qe} = \frac{\mathcal{C}_{s}}{\mathcal{C}_{\sigma}} T \sigma_{0} \Bigl\{ 1 + \frac{\Gamma_{imp}'}{\Gamma_{imp}} + \mathcal{C}_{s}^{-1} \Bigl(\frac{e}{c}\Bigr)^{2} \Bigl( \int \frac{d^{3} \bm{p}}{(2\pi)^{3}} \frac{\epsilon_{p} - \mu}{T} \Bigl( - \frac{\partial f_{eq}(\epsilon_{p})}{\partial \epsilon_{p}} \Bigr) [\boldsymbol{\Omega}_{\boldsymbol{p}} \cdot \bm{\hat{n}}_{\bm{p}}]^{2} \Bigr) B^{2} \Bigr\} , \nn && L_{xx}^{qq} = \frac{\mathcal{C}_{\kappa}}{\mathcal{C}_{\sigma}} T^{2} \sigma_{0} \Bigl\{ 1 + \frac{\Gamma_{imp}'}{\Gamma_{imp}} + \mathcal{C}_{\kappa}^{-1} \Bigl(\frac{e}{c}\Bigr)^{2} \Bigl( \int \frac{d^{3} \bm{p}}{(2\pi)^{3}} \frac{(\epsilon_{p} - \mu)^{2}}{T^{2}} \Bigl( - \frac{\partial f_{eq}(\epsilon_{p})}{\partial \epsilon_{p}} \Bigr) [\boldsymbol{\Omega}_{\boldsymbol{p}} \cdot \bm{\hat{n}}_{\bm{p}}]^{2} \Bigr) B^{2} \Bigr\} , \eqa
\end{widetext}
where contributions of both chiral Fermi surfaces are summed. $\mu^{+} = \mu^{-} = \mu$ being considered, i.e., $\mu^{5} = \mu^{+} - \mu^{-} = 0$, both the chiral magneto-electric effect and chiral magneto-heat effect are canceled by a pair of chiral Fermi surfaces. The Berry magnetic field is given by $\bm{\nabla}_{\bm{p}} \cdot \bm{\Omega}_{\bm{p}} = \delta^{(3)}(\bm{p})$ in a transformed coordinate and $\bm{\hat{n}}_{\bm{p}}$ is a unit vector defined on one chiral Fermi surface, associated with the Fermi velocity. $\sigma_{0} = \mathcal{C}_{\sigma} e^{2} N_{F} v_{F}^{2} \Gamma_{imp}^{-1}$ is a residual electrical conductivity with a positive numerical constant $\mathcal{C}_{\sigma}$. $\mathcal{C}_{s}$ and $\mathcal{C}_{\kappa}$ are also positive numerical constants for Seebeck and thermal-conductivity coefficients. Of course, a finite Seebeck coefficient requires particle-hole symmetry breaking near the chiral Fermi surface \cite{Mahan_Book}.


Inserting these expressions into Eq. (2), we find all transport coefficients,
\begin{widetext}
\bqa && \sigma_{xx} = \sigma_{0} \Bigl\{ 1 + \frac{\Gamma_{imp}'}{\Gamma_{imp}} + \Bigl(\frac{e}{c}\Bigr)^{2} Q_{\sigma} B^{2} \Bigr\} , ~~~~~ s_{xx} = - \frac{\mathcal{C}_{s}}{\mathcal{C}_{\sigma}} \frac{1 + \Gamma_{imp}'/\Gamma_{imp} + \mathcal{C}_{s}^{-1} (e/c)^{2} Q_{s} B^{2}}{1 + \Gamma_{imp}'/\Gamma_{imp} + (e/c)^{2} Q_{\sigma} B^{2}} , \nn && \frac{\kappa_{xx}}{T} = \sigma_{0} \Bigl[ \frac{\mathcal{C}_{\kappa}}{\mathcal{C}_{\sigma}} \Bigl\{ 1 + \frac{\Gamma_{imp}'}{\Gamma_{imp}} + \mathcal{C}_{\kappa}^{-1} \Bigl(\frac{e}{c}\Bigr)^{2} Q_{\kappa} B^{2} \Bigr\} -  \frac{\mathcal{C}_{s}^{2}}{\mathcal{C}_{\sigma}^{2}} \frac{\bigl\{ 1 + \Gamma_{imp}'/\Gamma_{imp} + \mathcal{C}_{s}^{-1} (e/c)^{2} Q_{s} B^{2} \bigr\}^{2} }{1 + \Gamma_{imp}'/\Gamma_{imp} + (e/c)^{2} Q_{\sigma} B^{2}} \Bigr] , \eqa
\end{widetext}
where $Q_{\sigma}$, $Q_{s}$, and $Q_{\kappa}$ are apparently given by the momentum-integral expressions of Eq. (7) in front of $B^{2}$. It becomes clear that all longitudinal transport coefficients show positive magnetoconductivity as discussed before, where their enhancements are proportional to $B^{2}$. This anomaly-involved $B^{2}$ dependence results in violation of Wiedemann-Franz law, given by
\begin{widetext}
\bqa && L(B) = \frac{\mathcal{C}_{\kappa}}{\mathcal{C}_{\sigma}}\frac{1 + \frac{\Gamma_{imp}'}{\Gamma_{imp}} + \mathcal{C}_{\kappa}^{-1} \Bigl(\frac{e}{c}\Bigr)^{2} Q_{\kappa} B^{2}}{1 + \frac{\Gamma_{imp}'}{\Gamma_{imp}} + \Bigl(\frac{e}{c}\Bigr)^{2} Q_{\sigma} B^{2}} -  \frac{\mathcal{C}_{s}^{2}}{\mathcal{C}_{\sigma}^{2}} \frac{\bigl\{ 1 + \Gamma_{imp}'/\Gamma_{imp} + \mathcal{C}_{s}^{-1} (e/c)^{2} Q_{s} B^{2} \bigr\}^{2} }{\bigl\{1 + \Gamma_{imp}'/\Gamma_{imp} + (e/c)^{2} Q_{\sigma} B^{2}\bigr\}^{2}} \nn && \approx L_{0} + \Bigl(\frac{e}{c}\Bigr)^{2} \frac{1}{1 + \frac{\Gamma_{imp}'}{\Gamma_{imp}}} \frac{\mathcal{C}_{\sigma} Q_{\kappa} - ( \mathcal{C}_{\sigma} \mathcal{C}_{\kappa} - 2 \mathcal{C}_{s}^{2} ) Q_{\sigma} - 2 \mathcal{C}_{s} Q_{s}}{\mathcal{C}_{\sigma}^{2}} B^{2} , \eqa
\end{widetext}
where the Lorenz number is expressed as $L_{0} = \frac{\mathcal{C}_{\kappa}}{\mathcal{C}_{\sigma}} -  \frac{\mathcal{C}_{s}^{2}}{\mathcal{C}_{\sigma}^{2}} = \frac{\pi^{2}}{3} \Bigl(\frac{k_{B}}{e}\Bigr)^{2}$ with the Boltzmann constant $k_{B}$, set to be $1$. Since the Seebeck coefficient is determined by particle-hole symmetry breaking of the band structure, we are allowed to assume $\mathcal{C}_{\sigma},~\mathcal{C}_{\kappa} \gg \mathcal{C}_{s}$, finding $L(B) \approx L_{0} + \Bigl(\frac{e}{c}\Bigr)^{2} \frac{1}{1 + \frac{\Gamma_{imp}'}{\Gamma_{imp}}} \frac{ Q_{\kappa} - \mathcal{C}_{\kappa} Q_{\sigma}}{\mathcal{C}_{\sigma}} B^{2}$. Then, we obtain $\mbox{sgn}[L(B) - L_{0}] \approx \mbox{sgn}[\mathcal{C}_{s}^{-1} Q_{\kappa} - L_{0} Q_{\sigma}]$, which will be determined by details of band structures.

On the other hand, the anomaly-involved $B^{2}$ dependence disappears when currents are driven in perpendicular to the applied magnetic field, given by
\bqa && \sigma_{xx} = \Bigl( 1 + \frac{\Gamma_{imp}'}{\Gamma_{imp}} \Bigr) \frac{\sigma_{0}}{1 + \omega_{c}^{2} / \Gamma_{imp}^{2}} , ~~~~~ s_{xx} = - \frac{\mathcal{C}_{s}}{\mathcal{C}_{\sigma}} , \nn && \frac{\kappa_{xx}}{T} = \Bigl( \frac{\mathcal{C}_{\kappa}}{\mathcal{C}_{\sigma}} -  \frac{\mathcal{C}_{s}^{2}}{\mathcal{C}_{\sigma}^{2}} \Bigr) \Bigl( 1 + \frac{\Gamma_{imp}'}{\Gamma_{imp}} \Bigr) \frac{\sigma_{0}}{1 + \omega_{c}^{2} / \Gamma_{imp}^{2}} , \eqa where $\omega_{c} = \frac{e B}{m c}$ is the cyclotron frequency. In other words, effects of chiral Fermi surfaces do not exist in transverse transport coefficients. As a result, Wiedemann-Franz law holds.

One can generalize the present Boltzmann-equation framework, introducing weak anti-localization corrections into the Boltzmann equation phenomenologically \cite{KSKim_BE}. Indeed, we found the negative magnetoresistivity for electrical transport in the presence of weak anti-localization corrections, recently verified experimentally \cite{Kim_Review,Negative_LMR3}. Similarly, combined effects of both weak anti-localization and chiral anomaly on longitudinal thermal transport coefficients would be observed in Bi$_{1-x}$Sb$_{x}$ around $x = 3 \%$ under magnetic fields, where the role of weak anti-localization corrections in thermal conductivity is expected to be the same as that in electrical conductivity \cite{Weak_AL_Thermal_Conductivity1,Weak_AL_Thermal_Conductivity2,Weak_AL_Thermal_Conductivity3,Weak_AL_Thermal_Conductivity4,Weak_AL_Thermal_Conductivity5}.

In order to attribute the mechanism of anomalous thermal and thermoelectric transport properties to topological terms, we need to extend the electromagnetic axion term into a general covariant form on a curved manifold \cite{Chiral_Grv_Anomaly}. Indeed, a mixed gauge-gravity action in five dimensions with one auxiliary space has been shown to result in the mixed covariant anomaly in four dimensions, where each chiral current is not conserved by contributions from not only electromagnetic ``instantons" but also gravitational ``instantons" \cite{TLFL_EFT2}. This leads us to conclude that the underlying mechanism for negative thermal and Seebeck magneto-resistivities is chiral gravitational and U(1) mixed anomalies, respectively.

It is an open problem to evaluate longitudinal transport coefficients for Weyl metal in a diagrammatic way, where the role of chiral Fermi surfaces is not apparently introduced into longitudinal correlation functions \cite{Kim_Review}. On the other hand, transverse transport coefficients such as chiral magnetic effect and anomalous Hall effect have been shown to result from chiral anomaly and Berry curvature, respectively \cite{Chiral_Magnetic_Effect7,CME_Diagram1,CME_Diagram2}. We believe that it is essential to construct an effective field theory incorporating axion electrodynamics in an explicit way instead of the form of a topological term.

In summary, we found that axion electrodynamics of a pair of chiral Fermi surfaces gives rise to negative thermal and Seebeck ``resistivities" beyond negative magneto-electrical resistivity, which occur only when currents are driven along the direction of the applied magnetic field in Weyl metal. We attributed the mechanism of such anomalous thermal transport phenomena to chiral gravitational and U(1) mixed anomalies, respectively. In particular, we proposed a modified Wiedemann-Franz law in this topological Fermi-liquid state, where the Lorenz number has a correction from the chiral anomaly, proportional to distance-square of a pair of Weyl points, i.e., $B^{2}$.

This study was supported by the Ministry of Education, Science, and Technology (No. 2012R1A1B3000550 and No. 2011-0030785) of the National Research Foundation of Korea (NRF) and by TJ Park Science Fellowship of the POSCO TJ Park Foundation. K.-S. Kim appreciates fruitful discussions with Heon-Jung Kim, M. Sasaki, Jung-Hoon Han, and Heung-Sun Sim.


\begin{thebibliography}{9}
\bibitem{Comment_on_E} The applied electric field is defined as $E_{j} = E_{j}^{ext} - \frac{1}{e} \partial_{j} \mu$, containing the gradient of the chemical potential induced by the temperature gradient.
\end{thebibliography}
\end{document}